\documentclass[conference,a4paper]{IEEEtran}

\usepackage{mathtools}
\usepackage{float}
\usepackage{graphicx} 
\usepackage[colorinlistoftodos]{todonotes}
\graphicspath{ {images/} }

\usepackage[linesnumbered,ruled,vlined]{algorithm2e} 

\usepackage[utf8]{inputenc}
\usepackage[english]{babel}
\usepackage{amsthm,amssymb}

\usepackage{optidef}

\theoremstyle{definition}
\newtheorem{lemma}{Lemma}
\newtheorem{corollary}{Corollary}

\newtheorem{definition}{Definition}
\newtheorem{example}{Example}
\newtheorem{remark}{Remark}

\makeatletter
\renewcommand{\fnum@figure}{Fig. \thefigure}
\makeatother

\DeclareMathOperator{\sgn}{sgn}
\DeclareMathOperator{\bin}{bin}
\DeclareMathOperator{\ffo}{ffo}
\DeclareMathOperator{\ffz}{ffz}
\DeclareMathOperator{\flz}{flz}

\IEEEoverridecommandlockouts                              
\title{Efficient Partial Rewind of \\Successive Cancellation-based Decoders
}

\author{
\IEEEauthorblockN{Mohammad Rowshan, \textit{Student Member, IEEE} and Emanuele Viterbo, \textit{Fellow, IEEE}}
 \thanks{This research work is supported by the Australian Research Council under Discovery Project ARC DP160100528.}\\
  \thanks{The authors are with the department of electrical and computer systems engineering (ECSE), Monash University, Melbourne, VIC 3800, Australia. (email: mohammad.rowshan@monash.edu, emanuele.viterbo@monash.edu)}\\
}

\begin{document}

\maketitle
\thispagestyle{empty}
\pagestyle{empty}

\begin{abstract}
Successive cancellation (SC) process is an essential component of various decoding algorithms used for polar codes and their variants. Rewinding this process seems trivial if we have access to all intermediate log-likelihood ratios (LLRs) and partial sums. However, as the block length increases, retaining all of the intermediate information becomes inefficient and impractical. Rewinding the SC process in a memory-efficient way is a problem that we address in this paper. We first explore the properties of the SC process based on the binary representation of the bit indices by introducing a new operator used for grouping the bit indices. This special grouping helps us in finding the closest bit index to the target index for rewinding. We also analytically prove that this approach gives access to the untouched intermediate information stored in the memory which is essential in resuming the SC process.  Then, we adapt the proposed approach to multiple rewinds, and apply it on SC-flip decoding and shifted-pruning based list decoding. The numerical evaluation of the proposed solution shows a significant reduction of $\geq 50\%$ in the complexity of the additional decoding attempts at medium and high SNR regimes for SC-flip decoding and less for shifted-pruning based list decoding.
\end{abstract}



\section{INTRODUCTION}
Polar codes \cite{arikan} are the first class of constructive channel codes that was proven to achieve the symmetric (Shannon) capacity of a binary-input discrete memoryless channel (BI-DMC) using a low-complexity successive cancellation (SC) decoder. 
Decoding of polar codes and their variants requires passing the channel log-likelihood ratios (LLRs) through a factor graph shown in \ref{fig:eff_llrs}. 
The evolved information at the output of the factor graph is used to make a hard decision or to calculate a metric in the SC-based decoders. The evolved LLR, a.k.a decision LLR, is obtained for each bit-channel successively. To calculate each decision LLR, we need to access the intermediate information on the factor graph. There are two ways to access them: 1) We can store all $N\cdot\log_2 N$ intermediate values, including LLRs and partial sums on the factor graph. This approach is acceptable for short codes under SC decoder or Fano decoder. However, as the code gets longer, in particular under list decoding or stack decoding, this approach will be expensive in terms of memory requirement. 2) We can store a portion of the intermediate values. It was observed in \cite{leroux} that for calculating every decision LLR, we need at most $N-1$ intermediate LLRs (excluding channel LLRs) and $N-1$ partial sums at any decoding step.

Some decoding schemes rely on additional decoding attempts when the decoding process fails in the first attempt. These schemes are as follows:
    1) SC-flip decoding: In this scheme, when SC decoding fails, the decoding is repeated from scratch, while in the additional attempts, the value of a single or multiple bits are flipped throughout the SC decoding process to correct the error caused by the channel noise and avoid propagation of this error  \cite{afisiadis}.
    2) Shifted-pruning based list  decoding: In this scheme \cite{rowshan-sp,rowshan-sp2,lv}, when SC list decoding fails, additional decoding attempts may correct the error given we shift the path pruning window at the position where the correct path was pruned from the list in the first decoding attempt. Note that a special case of this scheme is also called SCL flip scheme, bit-flipping for SCL, or by other names.
    3) Fano decoding: In this scheme \cite{rowshan-pac,rowshan-fano}, the decoder may have a back-tracking or backward movement to explore the other paths on the decoding tree. Unlike the first two schemes where the decoding of a codeword is completed, and then the additional decoding is repeated from the first bit, in the Fano algorithm, the backward movement occurs frequently somewhere between the first bit and the last bit. It might be better to use $N\log N$ memory for intermediate information in Fano decoding of very short codes rather than $N-1$ memory elements. This way, we the complexity reduces significantly at the cost of a larger memory requirement.

In \cite{rowshan-pac} and \cite{rowshan-fano}, we proposed a sophisticated algorithm to do the partial rewinding in Fano decoding. In this work, we propose a simple analytically-proved approach to efficiently rewind the SC algorithm to the position that we need to flip the value of a bit in SC-flip decoding or to shift the pruning window in the shifted-pruning scheme. This approach is designed based on the scheduling properties of the SC process and an operator that we introduce in this paper. The approach relies on a special grouping of the bit indices in $[0,N-1]$ based on the introduced operator. We also prove that the suggested bit index to resume the SC process utilizes the untouched intermediate information in the memory left from the previous decoding stages. The grouping of bit indices eases the job of finding the closest bit index to the target bit index for rewinding. This closeness contributes to the significant reduction of the time and computational complexity of the underlying decoding scheme while it does retain the error correction performance. We also adapt the proposed approach for multiple rewinds, which is a bit different with a single rewind, and we apply it on SC-flip decoding and shifted-pruning scheme for list decoding.

\textbf{Paper Outline:}
The rest of the paper is organized as follows. Section II introduces the notation for the polar codes and describes the intermediate information of the SC process. Section III review the details of updating schedule for intermediate information based on the binary representation of the bit indices. In Section IV, the properties of the SC process by introducing an operator and a special grouping scheme are explored. Section V proposes a simple approach for single and multiple rewinds of the SC process. In Section VI, we evaluate the complexity reduction of the proposed approach by applying it on the SC-flip and shifted-pruning sachems. Finally, Section VII makes concluding remarks.

\section{PRELIMINARIES}

A polar code of length $N = 2^n$ with $K$ information bits is constructed by choosing $K$ good bit-channels in the polarized vector channel for transmitting the information bits and optional auxiliary CRC or parity bits. The indices of these bit-channels are collected in the set $\mathcal{A}$. The rest of the $N-K$ bit-channels are used for transmitting known values as redundancy. Polar codes are encoded by $x_0^{N-1}\!=\!u_0^{N-1}G_N$ where $u^{N-1}_0\!=\!(u_0, . . . , u_{N-1} )$ is the input vector, and $G_N = B_NG_2^{\otimes n}$, where
$G_2 \overset{\Delta}{=}{\footnotesize \begin{bmatrix}
1 & 0 \\
1 & 1
\end{bmatrix} }$, $B_N$ is an $N \times N$ bit-reversal permutation matrix, and $(\cdot)^{\otimes n}$ denotes the $n$-th Kronecker power \cite{arikan}. 
Let $y^{N-1}_0=(y_0, . . . , y_{N-1})$ denote the output vector of a noisy channel, and $\mathbf{\lambda}=(\lambda_0, . . . , \lambda_{N-2})$ vector indicates the long likelihood ratios (LLRs). The channel LLRs are computed based on the received signals from the physical channel, $y^{N-1}_0$. 

We also have intermediate LLRs as shown in Fig. \ref{fig:eff_llrs}. The intermediate LLRs are computed based on the type of node in the factor graph. $f$ and $g$ nodes are shown by circles and rectangles, respectively, in the factor graph. The output of these nodes can be computed from right to left by
\begin{equation}
\label{eq:f_func}
 f(\lambda_a,\lambda_b)\approx\sgn(\lambda_a)\cdot \sgn(\lambda_b)\cdot \min(|\lambda_a|,|\lambda_b|)
\end{equation}
\begin{equation}
\label{eq:g_func}
 g(\lambda_a,\lambda_b,\hat{\beta})=(-1)^{\hat{\beta}}\lambda_a+\lambda_b
\end{equation}
where $\lambda_a$ and $\lambda_b$ are the input LLRs to a node and $\hat{\beta}$ is the partial sum of previously decided bits corresponding to feed the estimated bits $\hat{u}_i$ backward into the factor graph.





In the SC decoding, the non-frozen bits are estimated successively based on the evolved LLRs via a one-time pass through the factor graph. 
When decoding the $i$-th bit, if $i \notin \mathcal{A}$, then $\hat{u}_i=0$ since $u_i$ is a frozen bit. Otherwise, bit $u_i$ is decided by a maximum likelihood (ML) rule 
$h(\lambda_0)$. Unlike the SC decoding which makes a final decision for $i$-th bit, SC list decoding considers both possible values $u_i= 0$ and $u_i=1$. In SC list decoding, the $L$ most reliable paths are preserved at each decoding step to limit growing of the number of paths. The solution for decoding is chosen at the lest bit based on the likelihood or the cyclic redundancy check (CRC) approach. The cyclic redundancy check (CRC) is also used in the re-decoding schemes such as SC-flip and the shifted-pruning based list decoding.

\section{Updating the Intermediate Information}\label{sec:eff_memory}
\subsection{Intermediate LLRs}
\begin{figure}[t]
    \centering
    \includegraphics[width=1\columnwidth]{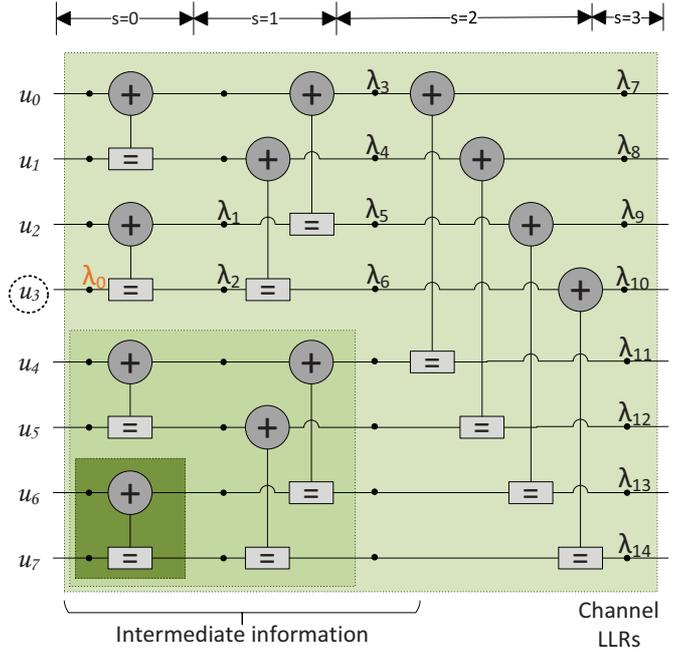}
    \caption{An illustrative example for updating LLRs for decoding bit $u_3$. $\lambda_0$ is computed based on $\lambda_1$, $\lambda_2$ and $\beta_0=\hat{u}_2$ (see Fig. \ref{fig:eff_ps}).}
    \label{fig:eff_llrs}
    \vspace{-5pt}
\end{figure}

The factor graph shown in Fig. \ref{fig:eff_llrs} has $N\log_2 N$ nodes however, as it was shown in \cite{leroux}, it is sufficient to update/access at most $N-1$ intermediate LLRs  out of $N\log_2 N$ LLRs for decoding any bit $u_i, 0\leq i\leq N-1$. Fig. \ref{fig:eff_llrs} illustrates the LLRs associated with decoding bit $u_3$ in a tree form on the factor graph. 
As can be seen, there are $2^s$ LLRs in stage $s$ for $s=0,...,n$. Hence, according to the geometric series, we need a total memory space of
\begin{equation}
    \sum_{s=0}^n2^s=2^{n+1}-1=2N-1
\end{equation}

Suppose $u_i$ is the bit that was just decoded and $\bin(i)=i_{n-1}...i_0$ is the binary representation of index $i$ where the least significant bit is indexed 0 and most significant bit is indexed $n-1$. . The stages are updated from right to left (where $s=0$). The first stage to be updated is obtained by {\em finding the first one, ffo}, or the position of the least significant bit set to one as
\begin{equation}\label{eq:eta}
    \eta(i)=\ffo(i_{n-1}...i_0)=\begin{dcases}
 \underset{i_t=1}{\min}(t) & i>0\\ n-1 & i=0
\end{dcases}    .
\end{equation}

Note that in the semi-parallel hardware architecture \cite{leroux}, since the LLRs are stored in blocks, memory usage is\,inefficient such that there will be some unused memory space. In fact, the reduction in the number of processing elements is traded with slightly higher clock cycles and larger memory space.

\subsection{Partial Sums}
The Partial sums are the other set of intermediate information needed for the SC process. It turns out that we need the same memory space for the partial sums as well, i.e., at most $N-1$ memory elements. It was observed in \cite{leroux} that we need to store $2^s$ bits corresponding to $2^s$ nodes of type $\mathbf{g}$ at stage $s$, which are waiting to be summed with the next decoded bit. Here, let us define an operator that indicates the last stage to be updated. The last stage that its partial sums to be updated is obtained by {\em finding the first zero, ffz}, or the position of the least significant bit set to zero as
\begin{equation}
    \psi(i)=\ffz(i_{n-1}...i_0)=\underset{i_t=0}{\min}(t).
\end{equation}
It turns out that this is the only stage that consists of $\mathbf{G}$ nodes in the process of updating LLRs from stage $s=\ffo(\bin(i))$ up to $s=0$. Clearly, after decoding the last bit where there is no zero in the binary representation of the index, $\bin(N-1)=11...1$, there is no need to update partial sums.

\begin{figure}[t]
    \centering
    \includegraphics[width=1\columnwidth]{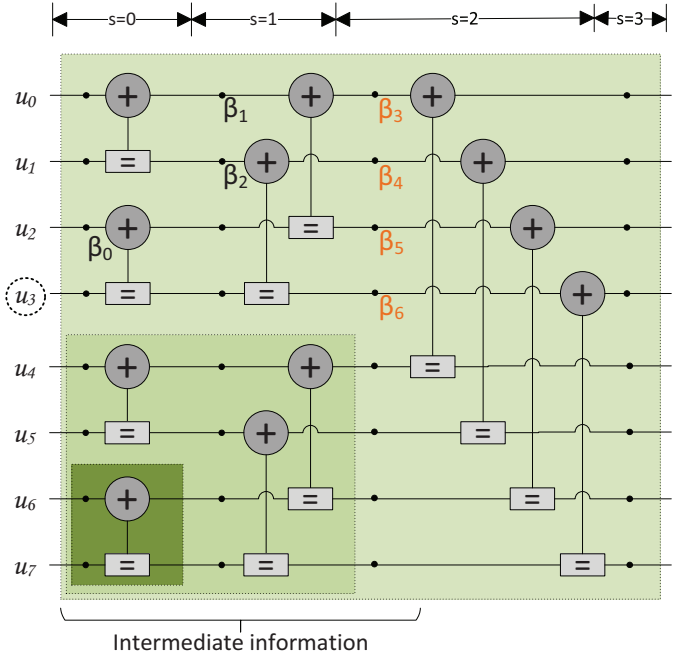}
    \caption{An illustrative example for updating partial sums of stage $s=2$ after decoding $u_3$.}
    \label{fig:eff_ps}
    \vspace{-5pt}
\end{figure}

Fig. \ref{fig:eff_ps} shows $N-1=8-1=7$ partial sums ($\beta_0$ to $\beta_6$) associated to $u_3$. The $\beta$ values in orange are updated after decoding bit $u_3$ as $\beta_6=u_3$, $\beta_5=u_3\oplus \beta_0$, $\beta_4=u_3\oplus \beta_2$, and $\beta_3=u_3\oplus \beta_1\oplus\beta_3$.

There are methods proposed in \cite{berhault,fan} for hardware implementation that require slightly less memory space for updating the partial sums.

You may notice that for $i\in[0,N-2]$, we have
\begin{equation}\label{eq:psi_i=eta_i+1}
    \psi(i)=\eta(i+1)
\end{equation}
That is the reason why at any bit $i\in[1,N-1]$, the stage $\eta(i)$ where  its LLRs needs to be updated consists of only $\mathbf{g}$ nodes. Therefore, after decoding bit $i-1$, the partial sums  of this stage are updated to be used for the $\mathbf{g}$ nodes at stage $\eta(i)$.
 
\section{Properties of the SC Process}\label{sec:SC_properties}
We discover some properties of the SC algorithm that can help us to rewind the process efficiently. The goal is not storing all the $N\log_2 N$ values for LLRs and partial sums or restarting the SC process from bit 0 in the SC-based decoding when a re-decoding attempt is required. 
First, let us define an operator that helps us in the upcoming analysis.
\begin{definition}\label{def:phi}
The operator $\phi(j)$ finds the last zero, flz, or the position of the most significant bit set to zero in the binary representation of $j=(j_{n-1}...j_0)_2$ indexed in reverse order as 
\begin{equation}
    \phi(j)=\flz(j_{n-1}...j_0)=\begin{dcases}
    n-1-\underset{j_t=0}{\max}(t) & j<2^n-1,\\
    n-1 & j=2^n-1 
    \end{dcases}
\end{equation}
for every $t\in[0,n-1]$. 
We denote the output of the operator $\phi(j)$ by parameter $p$.
\end{definition}
Note that since the indexing is in the opposite direction when the most significant bit is set to zero, i.e., $j_{n-1}=0$, then we get $p=0$, and when the only zero bit is $j_0=0$ or there is no 0-value bit, then $p=n-1$.
\begin{definition}[Set $\mathcal{Z}_p$]\label{def:Zp}
We group the bit indices $j\in[0,1,...,2^n-1]$ based on the identical $p=\phi(j)$ into $n$ sets denoted by set $\mathcal{Z}_p$ with order $p=0,1,...,n-1$, or
\begin{equation}
    \mathcal{Z}_p = \{ j\in[0,2^{n}-1] : \phi(j)=p  \}
\end{equation}
\end{definition}

\begin{example}
For $n=3$, we can group the indices 0 to 7 into the following sets: 
$\mathcal{Z}_0=\{0,1,2,3\}$, $\mathcal{Z}_1=\{4,5\}$, and  $\mathcal{Z}_2=\{6,7\}$.
\end{example}
\begin{remark}\label{rmk:nfz_bits_distrib}
    The distribution of non-frozen indices in set $\mathcal{A}$ among sets $\mathcal{Z}_p,p\in[0,n-1]$ depends on the code rate. As the code rate reduces, a fewer non-frozen indices will exist in low order  $\mathcal{Z}_p$, i.e., $\mathcal{Z}_p$ with small $p$.
\end{remark}

\begin{lemma}[Properties of $\mathcal{Z}_p$]\label{lma:Zp_properties}
For any $n>0$ and $p\in[0,n-1]$, set $\mathcal{Z}_p$ has the following properties:
\begin{enumerate}
\item[a.] The boundaries of set $\mathcal{Z}_p$ are
\begin{equation}
    \mathcal{Z}_p = \begin{dcases}
    [2^{n}\!-\!2^{n-p}, 2^{n}\!-\!2^{n-(p+1)}\!-\!1] & 0\leq p<n-1,\\
    [2^{n}\!-\!2^{n-p}, 2^{n}\!-\!1] & p=n-1
    \end{dcases}
\end{equation}
\item[b.] The size of set $\mathcal{Z}_p$ is 
\begin{equation}
    |\mathcal{Z}_p|=\begin{dcases}
    2^{n-p-1} & 0\leq p<n-1,\\
    2 & p=n-1
    \end{dcases}
\end{equation}
\item[c.] The smallest element in set $\mathcal{Z}_p$ is
\begin{equation}
    z_p = \min(\mathcal{Z}_p)=\sum_{x=n-p}^{n-1} 2^x = 2^n-2^{n-p}
\end{equation}
\end{enumerate}
\end{lemma}
\begin{proof}
Let us first introduce a notation for the binary representation of a positive integer with length $n$. Given 
$\{0,1\}^x$ indicates a mixed string of 0 and 1, and $\{b\}^x, b\in\{0,1\}$ denotes a uniform string of either 0 or 1, both  with length $x$. In set $\mathcal{Z}_p, p<n-1$, observe that the elements are in the form of $\{1\}^p+\{0\}+\{0,1\}^{n-(p+1)}$ where the operator '+' is used for concatenation and $\{1\}^p$ is/are the most significant bits. 
\begin{enumerate}
\item[a.] 
The smallest element of set $\mathcal{Z}_p$ in binary is
$$\{1\}^p+\{0\}+\{0\}^{n-(p+1)}=\{1\}^p+\{0\}^{n-p}$$
which is equivalent to 
$$\sum_{x=n-p}^{n-1} 2^x=2^n-2^{n-p}$$ 
in decimal. Similarly, one can see that the largest element in set $\mathcal{Z}_p, p<n-1$ is $$\{1\}^p+\{0\}+\{1\}^{n-(p+2)}=\big(\{1\}^n\big)_2-\big(\{1\}+\{0\}^{n-(p+2)}\big)_2$$  
which is equivalent to $(2^n-1)-2^{n-(p+1)}$ in decimal.

Note that the largest element in set $\mathcal{Z}_p, p=n-1$ is $\big(\{1\}^n\big)_2=2^{n-1}$ while the smallest element follows the relationship discussed above.

\item[b.] Given the interval $[\min(\mathcal{Z}_p), \max(\mathcal{Z}_p)]$ in part a of this lemma, we can find the size of set $\mathcal{Z}_p$ by $\max(\mathcal{Z}_p)-\min(\mathcal{Z}_p)+1$.

\item[c.] It follows from part a of this lemma that the lower bound of the values in set $\mathcal{Z}_p$ in binary is
$\{1\}^p+\{0\}^{n-p}$ which is equivalent to $\sum_{x=n-p}^{n-1} 2^x=2^n-2^{n-p}$ in decimal.
\end{enumerate}
\end{proof}
\begin{example}
For $n=4$, we have $z_p$ for $p=0,1,...,n-1$ as
$$z_0=(0000)_2=0, z_1=(1000)_2=8$$ $$z_2=(1100)_2=12, \text{ and }z_3=(1110)_2=14$$
or based on the lower bound of $\mathcal{Z}_p$ in Lemma \ref{lma:Zp_properties} as
$$z_0=2^n-2^{n-0}=16-2^4=0, z_1=16-2^3=8$$ $$z_2=16-2^2=12, \text{ and }z_3=16-2=14$$
\end{example}

Let us find the deepest updated stage while decoding any bit $i$ within set $\mathcal{Z}_{p}$ in the following lemma. 
\begin{lemma}\label{lma:max_eta}
For any $i\in\mathcal{Z}_p,p\in[0,n-1]$, and $z_p=\min(\mathcal{Z}_p)$ we have
\begin{equation} 
    \underset{i\in\mathcal{Z}_p}{\max}(\eta(i))=\eta(z_p)
\end{equation}
\end{lemma}
\begin{proof}
Let us recall the notation $\{1\}^p+\{0\}+\{0,1\}^{n-(p+1)}$ for $i\in\mathcal{Z}_p,p<n-1$ where the operator '+' is used for concatenation and $\{1\}^p$ is/are the most significant bits. According to \eqref{eq:eta}, the maximum value for $\eta(i)$, i.e., the largest index for the least significant bit set to one for $i\in\mathcal{Z}_p$, is obtained when we have $$\bin(i)=\{1\}^p+\{0\}+\{0\}^{n-(p+1)}=\{1\}^p+\{0\}^{n-p}$$
which is the smallest element in set $i\in\mathcal{Z}_p,p<n-1$, i.e., $z_p=\min(\mathcal{Z}_p)$. 

For $p=n-1$,  although the notation is in the form of $\{1\}^{n-1}+\{0,1\}^1$, the largest index for the least significant bit set to one is similarly obtained from $\{1\}^{n-1}+\{0\}$ which is the smallest in set $\mathcal{Z}_{n-1}$
\end{proof}
Clearly, when $p=0$, we have $\max(\eta(i))=\eta(0)=n-1$ for any $i\in\mathcal{Z}_0$.

\begin{remark}\label{rmk:update_p}
From Lemma \ref{lma:max_eta} we conclude that the deepest  stage that the intermediate LLRs are overwritten/updated is when decoding the smallest bit index in set $\mathcal{Z}_{p}$. Recall that the partial sums used at stage $\eta(z_p)$ are provided after decoding bit with index $z_p-1$ according to \eqref{eq:psi_i=eta_i+1}.
\end{remark}

Now we consider updating intermediate information for $i$ in different sets of  $\mathcal{Z}_{p}$.
\begin{lemma}\label{lma:eta_z_p_p'}
For any $p,p'\in[0,n-1]$, $p<p'$, we have
\begin{equation}
    \eta(z_{p}) > \eta(z_{p'})
\end{equation}
\end{lemma}
\begin{proof}
It follows directly from \eqref{eq:eta} and part c of Lemma \ref{lma:Zp_properties}. Note that $z_p$ is in the form of $\{1\}^p+\{0\}^{n-p}$. It can be observed that for the smaller $p$, the position of the least significant bit set to one has a larger index. Therefore, $\eta(z_{p})$ is larger. 
\end{proof}
\begin{corollary}\label{cor:psi_z_p_p'}
For any $p,p'\in[0,n-1]$,  $p<p'$, we have
\begin{equation}
    \psi(z_{p}-1) > \psi(z_{p'}-1)
\end{equation}
\end{corollary}
\begin{proof}
As $\eta(z_p)=\psi(z_p-1)$ according to \eqref{eq:psi_i=eta_i+1}, then based on Lemma \ref{lma:eta_z_p_p'}, it follows that $\psi(z_{p}-1) > \psi(z_{p'}-1)$. 
\end{proof}

\begin{remark}\label{rmk:update_p_p'}
From Lemma \ref{lma:eta_z_p_p'} and Corollary \ref{cor:psi_z_p_p'}, we conclude that intermediate LLRs and partial sums of stage $\eta(z_{p})$ are not overwritten/updated when we are decoding any bit with index $i\in\mathcal{Z}_{p'}, p<p'$. 
\end{remark}
\begin{remark}\label{rmk:no_update_rewined}
As per Remark \ref{rmk:update_p_p'} and the fact that updating the intermediate information is performed from stage $\eta(z_{p})$ to stage 0, rewinding the SC algorithm from bit $i\in\mathcal{Z}_{p'}$ to bit $z_p$, $p<p'$ does not require any additional update of the intermediate LLRs or partial sums. 
\end{remark}
We will use remarks \ref{rmk:update_p_p'} and \ref{rmk:no_update_rewined} in the proposed approach later.






\section{Efficient Partial Rewind}\label{sec:rewinding}
We learned in Section \ref{sec:eff_memory} that we could save memory significantly by knowing the required intermediate LLRs and partial sums needed for decoding each bit. However, there is a drawback to this efficiency. Since we use limited space for intermediate information instead of $N\log_2 N$ memory elements, we have to overwrite the current values we no longer need to proceed with decoding. In the normal decoding process, the overwriting operation does not cause any data corruption. However, if we need to move backward like in SC-flip, shifted-pruning, or Fano decoding, we may no longer access the intermediate information as it may have been lost due to overwriting. 

In this section, based on the properties of the SC process we studied in Section \ref{sec:SC_properties}, a scheme is proposed such that rewinding the SC algorithm is performed efficiently by significantly fewer computations comparing with restarting the algorithm.

Suppose the SC algorithm is decoding bit $i$ and needs to rewind the SC process to bit $j, j<i$, and $i,j\in\mathcal{A}$. In SC-flip scheme and shifted-pruning-scheme, we have $i=2^n-1$ however, in Fano decoding, $i\leq2^n-1, i\in\mathcal{A}$. Since the required intermediate information for decoding bit $j$ may partially be overwritten, we may need to rewind further to a position denoted by $j_p$. From $j_p$, the SC algorithm proceeds with the normal decoding up to position $j$. We shift the pruning window at this position, or we flip the bit $u_j$ and then continue the normal SC-based decoding.

Now, the question is what the position $j_p$ is? Let us assume $i\in\mathcal{Z}_{p'}$ and $j\in\mathcal{Z}_{p}$. Then, 
\begin{equation}\label{eq:j_p}
j_p=
    \begin{dcases}
 z_p & \text{if } z_p<z_{p'}\\ 
 z_{p'} & \text{if } z_p=z_{p'}
\end{dcases}
\end{equation}

\begin{example}
Suppose $N=2^5$ and we need to rewind the SC algorithm from position $i=31=(11111)_2$ to $j=19=(10011)_2$. We know that $i\in\mathcal{Z}_4$ and $j\in\mathcal{Z}_1$. Therefore, according to \eqref{eq:j_p},  $j_p=z_p=16=(10000)_2$.
\end{example}

\noindent{\bf Recursion for Case $\mathbf{z_p=z_{p'}}$}: For the case $z_p=z_{p'}$ in \eqref{eq:j_p}, we may choose a position $k$, $j_p<k\leq j$ for rewinding, which is more efficient. To this end, let us $k\gets j$ and $m\gets n$, then while $\phi(k)\neq 0$: \begin{itemize}
    \item first, truncate the binary representation of $k=(k_{n-1}...k_1k_0)_2$ by removing the bits from position $m-1-\phi(k)$ to the most significant bit (inclusive), i.e. position $m-1$. Note that after truncation, we have a binary number with length $m=m-(\phi(k)+1)$.
    \item secondly, find the new set $\mathcal{Z}_{p''}$ such that $k\in\mathcal{Z}_{p''}$ for $k=\sum_{t=m-(\phi(k)+1)}^{n-1}j_t\cdot2^{t}-k$.
    \item then, $j_{p}=\sum_{t=m-(\phi(k)+1)}^{n-1}j_t\cdot2^{t}+z_{p''}$.
\end{itemize}
We can continue the above procedure recursively to minimize $z_{p''}-j$. Note that in this recursion, $k$  and $m$ are being replaced with new values at each iteration.

\begin{example}
Suppose $N=2^5$ and we need to rewind the SC algorithm from position $i=22=(10110)_2$ to $j=19=(10011)_2$. We know that $i,j\in\mathcal{Z}_1$ and therefore $j_p=z_p=16=(10000)_2$. We truncate $j=(10011)_2$ as mentioned above. We get $k=(011)_2$, $k\in\mathcal{Z}_0$, and $z_{p''}=0$. Hence, the new $j_p$ is $j_p=z_p+z_{p''}=16$ which is the same as before.
\end{example}

\begin{example}
Suppose $N=2^5$ and we need to rewind the SC algorithm from position $i=22=(10110)_2$ to $j=20=(10100)_2$. We know that $i,j\in\mathcal{Z}_1$ and therefore $j_p=z_p=16=(10000)_2$. However, if we truncate $j=(10100)_2$ as mentioned above, we get $t=(100)_2$, $t\in\mathcal{Z}_1$, and $z_{p''}=4$. Hence, the new $j_p$ is $j_p=z_p+z_{p''}=16+4=20$.
\end{example}

One can observe that the recursion is not used in the schemes that the rewind is performed from the last bit index. The reason is that bit index $2^n-1\in\mathcal{Z}_n$ and this set has only one other element which is $2^n-2=z_{n-1}$. On the other hand, if we need to rewind the SC process to a bit index smaller than $z_{n-1}$, the target bit index will fall into another set $\mathcal{Z}_p$ with different $z_p$. Hence, this may be used for Fano decoding where the case $z_p=z_{p'}$ is possible. Note that we do not numerically evaluate this approach for Fano decoding as we do not have any other approach to compare with. We can either use this approach or simply we can store all $N\log_2 N$ intermediate LLRs and partial sums and trade a significant complexity reduction with the memory efficiency.

Now, let us adapt the proposed approach for rewinding more than once. In the shifted-pruning scheme (and in the SC-flip scheme), we may need to repeat the rewind of the SC process up to $T$ times. Therefore, we need to take this into our consideration. Assuming $t\in[1,T]$ indicates the current iteration, and $j(t)$ and $j_p(t)$ denotes the $j$ and $j_p$ of iteration $t$, then $j_p$ of the current iteration is obtained by considering $j_p(t-1)$ as follows:

\begin{equation}\label{eq:j_p_t}
j_p=
    \begin{dcases}
 j_p(t-1) & \text{if } j_p(t)>j_p(t-1)\\ 
 j_p(t) & \text{otherwise}
\end{dcases}
\end{equation}

As \eqref{eq:j_p_t} shows, if the destination position of the current iteration $j(t)$ is larger than the destination position of the previous iteration, the intermediate information is not valid. The reason is that some modification (bit-flipping or shifted-pruning) occurred at position $j(t-1)$ that affects not only the intermediate information but also the decoded data.  In other words, we need to go to position $j(t-1)$ and undo the modification and proceed with the decoding up to the position $j(t)$ and then perform the modification of the current iteration. Note that if both $j(t)$ and $j(t-1)$ are in the same $\mathcal{Z}_p$, then $j_p(t)=j_p(t-1)$, hence there will be no difference.

\begin{figure}
    \centering
    \includegraphics[width=1\columnwidth]{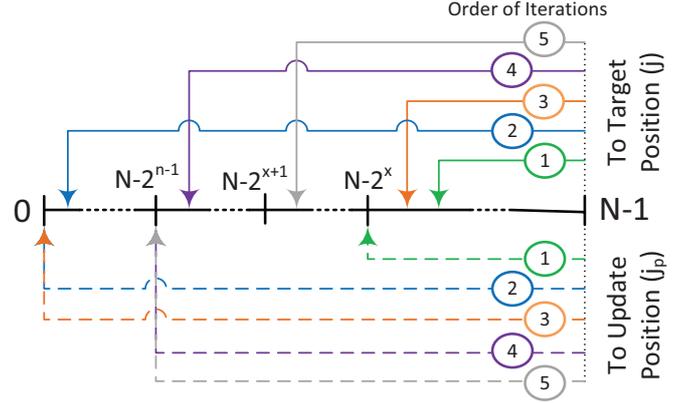}
    \caption{An illustrative example comparing the target position $j$ and update position $j_p$}
    \label{fig:eff_iter}
    \vspace{-5pt}
\end{figure}

Fig. \ref{fig:eff_iter} compares $j$ and $j_p$ for an example where 5 iterations are occurring.

Furthermore,  when rewinding the list decoder from the last bit position, $N-1$, to position $j_p$, some of the paths that existed at position $j_p$ in the previous iteration might be eliminated in between and be replaced with other paths. This potential replacement should be addressed when we have a list of paths/candidates, such as in the shifted-pruning scheme, not in the SC-flip scheme. 
To simplify the problem, we can limit the positions $j_p$ to $j_p=2^{n-1}$.  Because all the computations of the intermediate LLRs from this position, $2^{n-1}$, up to the last position, $2^n-1$, are performed solely based on the channel LLRs and partial sums of stage $\psi(2^{n-1}-1)$. Hence, we need to store the decoded data, $\mathbf{u}[0:2^{n-1}-1]$, and the path metric of all the paths at position $2^{n-1}-1$. Partials sums can be stored as well or can be computed simply by $\mathbf{u}[0:2^{n-1}-1]\mathbf{G}_{N/2}$.

\section{Numerical Results}\label{sec:rewinding_results}
We show that in the additional decoding attempts in SC list decoding and SC decoding, the average complexity (in terms of required time-steps and node visits) can be significantly reduced by partial rewinding instead of full rewinding of SC-based decoder. Note that taking average over all the decoding attempts including the successful attempts in the first run does not give a good insight and a fair comparison in particular at medium and high SNR regimes. The reason is that only a small portion of the total attempts fail requiring additional attempts, e.g., less than 10 failures in $10^4$ decoding attempts in the FER range of $10^-4$. Hence, the impact of this small portion becomes negligible on the average number of total attempts per codeword at high SNR regimes. 

\begin{figure}
    \centering
    \includegraphics[width=1\columnwidth]{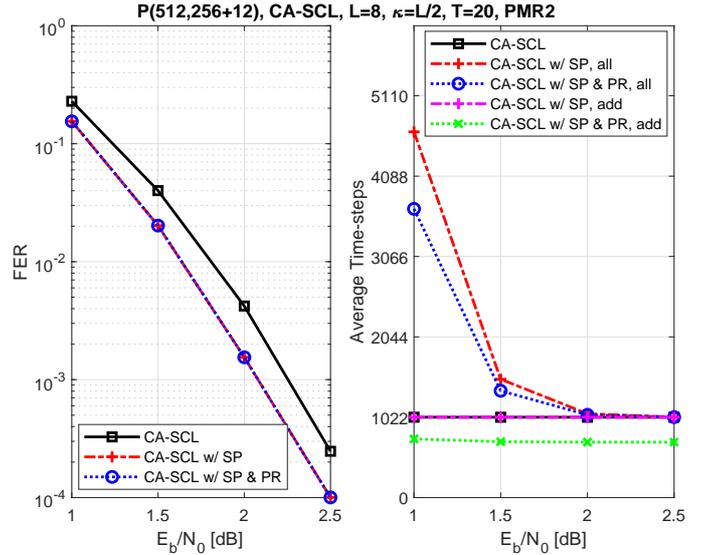}
    \caption{Comparison of FER and average time complexity of P(512,256+12) under CA-SCL decoding without and with (w/) shifted-pruning scheme (SP), and with partial rewinding (PR). 'all' and 'add' indicate average over all the decoding iterations and average only over additional iterations for shifted-pruning, respectively.}
    \label{fig:cplx_pr1}
    \vspace{-5pt}
\end{figure}

\begin{figure}
    \centering
    \includegraphics[width=1\columnwidth]{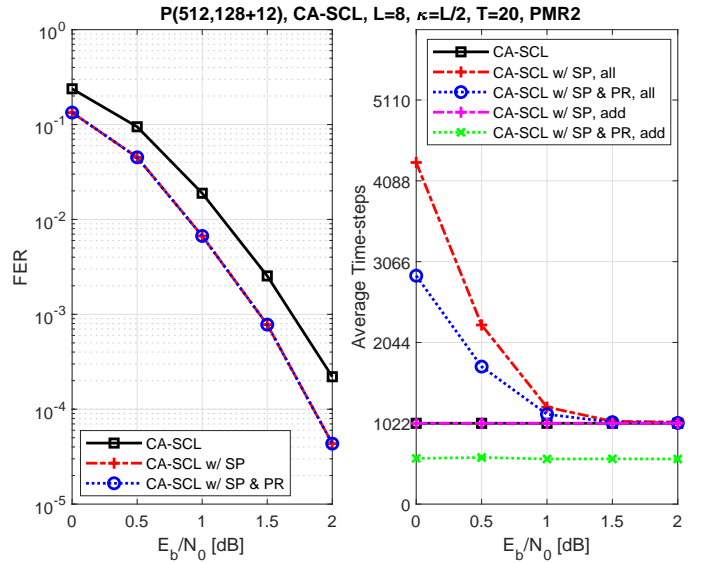}
    \caption{Comparison of FER and average time complexity of P(512,128+12) under CA-SCL decoding without and with (w/) shifted-pruning scheme (SP), and with partial rewinding (PR). 'all' and 'add' indicate average over all the decoding iterations and average only over additional iterations for shifted-pruning, respectively.}
    \label{fig:cplx_pr2}
    \vspace{-5pt}
\end{figure}

Figures \ref{fig:cplx_pr1} and \ref{fig:cplx_pr2} compare the average computational complexity of shifted-pruning scheme with and without partial rewinding for two different codes. In Fig. \ref{fig:cplx_pr1}, the FER and time complexity of polar code of (512,256+12) constructed with DEGA (2dB) \cite{trifonov} and concatenated with CRC12 with polynomial 0xC06 under SC list decoding with list size $L=8$ with shifted-pruning (SP) are shown. The FER before and after using the efficient partial rewind (PR) scheme clearly shows that the proposed efficient partial rewind scheme does not degrade the decoder's performance as we expected. However, it reduces the average time-steps over additional iterations (when the decoding fails) by over 30\% (from $2N-2=1022$ time-steps (or clock cycles) \cite{leroux} down to about 700 time-steps). The average time steps over all the iterations also reduce, but at high SNRs, it approaches 1022. The reason is that at high SNR regimes, the number of errors, FER, is low. Compared with the total number of codewords decoded successfully, just a small number of codewords are failed to be decoded in the first attempt and need additional attempts/iterations.

As Fig. \ref{fig:cplx_pr2} shows, the reduction in the average time complexity for efficient partial rewind scheme improves for polar code P(512,128+12) constructed with DEGA (1 dB). The average time-steps over additional iterations by about 45\% (from $2N-2=1022$ time-steps down to about 570 time-steps). The reason is that at a low code rate of $R=1/4$, the positions $j$ for shifting the pruning window are mostly located in the interval $[N/2,N-1]$ where the partial rewinding can be effective in reducing the complexity. Recall that if $j\in [0,N/2-1]=\mathcal{Z}_0$, then $j_p=z_p=0$. That means a full rewind is needed. One can guess that the reduction in the complexity would be less at high code rates where the position $j$ for shifting are dominantly located in $[0,N/2-1]=\mathcal{Z}_0$ as the reliability of these bit-positions are less relative to the ones in $[N/2,N-1]$.

\begin{figure}
    \centering
    \includegraphics[width=1\columnwidth]{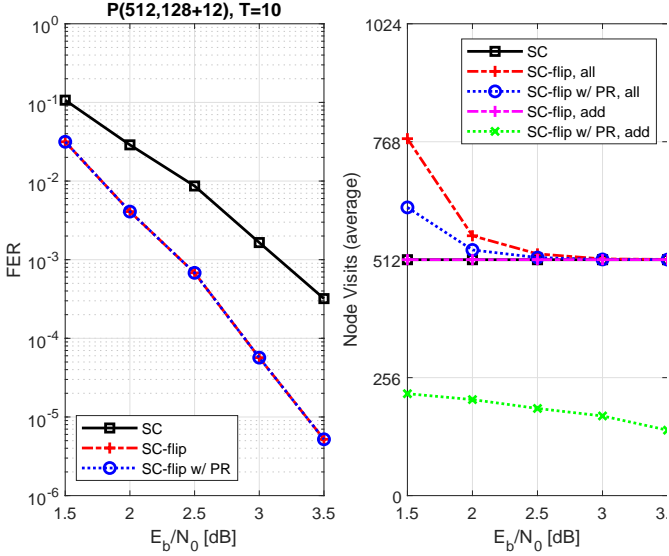}
    \caption{Comparison of FER and average node visits of P(512,128+12) under SC decoding without and with (w/) bit-flipping, and with partial rewinding (PR). 'all' and 'add' indicate averaging over all the decoding iterations and averaging only over additional iterations, respectively.}
    \label{fig:cplx_pr_flip1}
    \vspace{-5pt}
\end{figure}

\begin{figure}
    \centering
    \includegraphics[width=1\columnwidth]{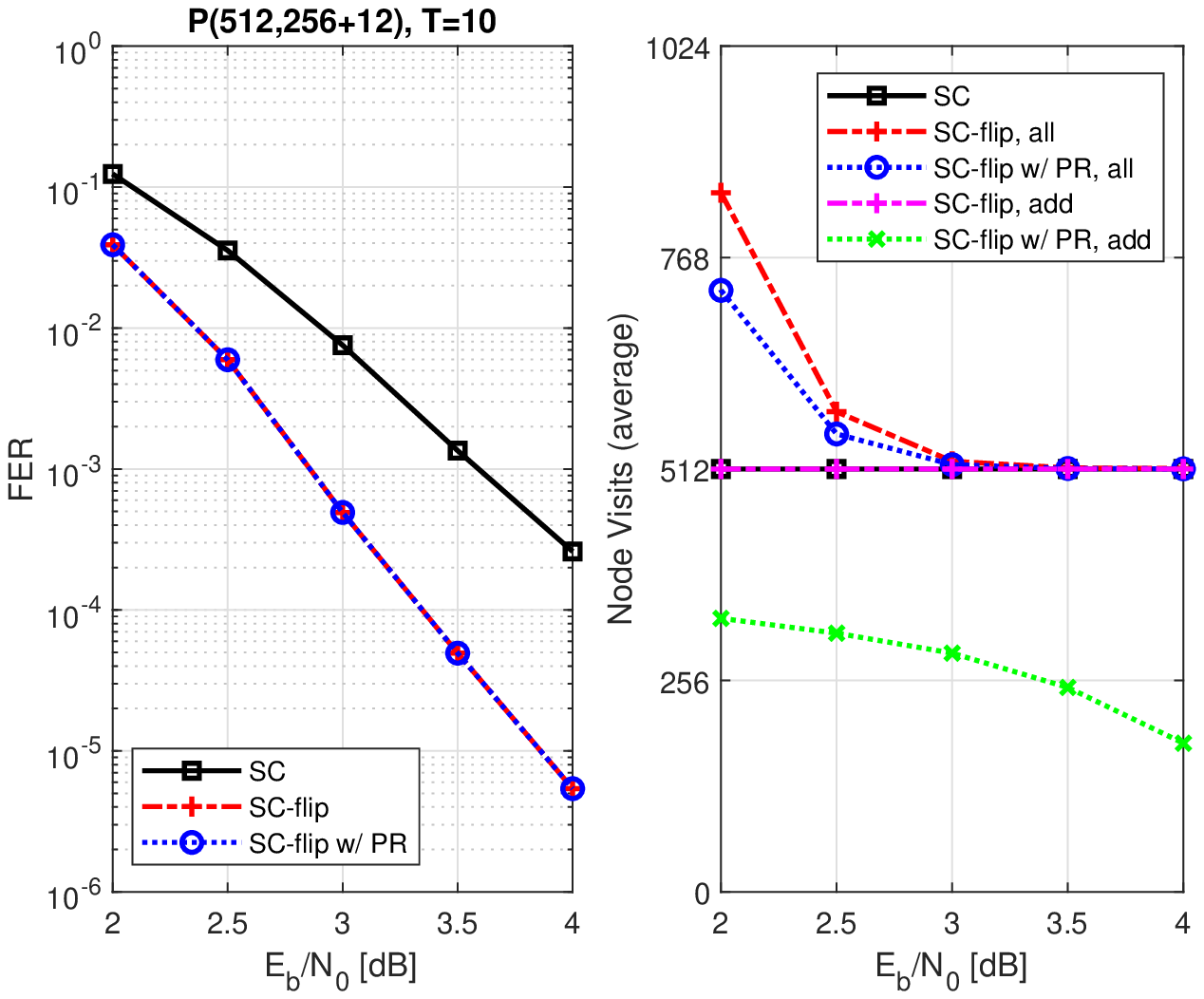}
    \caption{Comparison of FER and average node visits of P(512,256+12) under SC decoding without and with (w/) bit-flipping, and with partial rewinding (PR). 'all' and 'add' indicate averaging over all the decoding iterations and averaging only over additional iterations, respectively.}
    \label{fig:cplx_pr_flip2}
    \vspace{-5pt}
\end{figure}

\begin{figure}
    \centering
    \includegraphics[width=1\columnwidth]{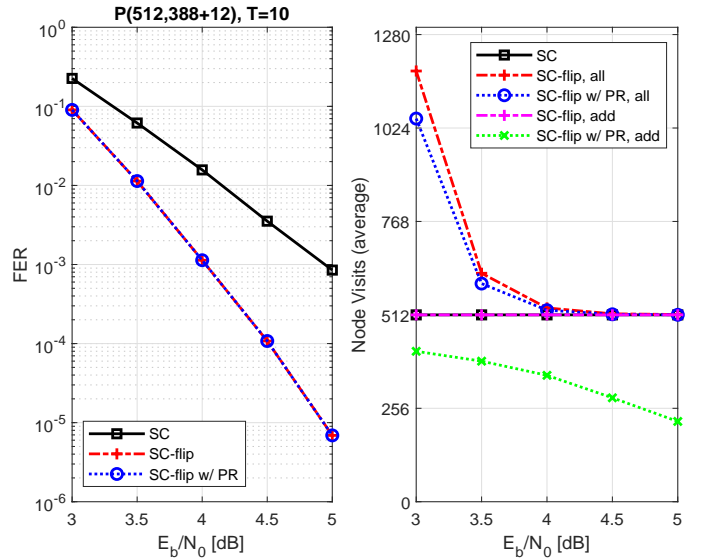}
    \caption{Comparison of FER and average node visits of P(512,388+12) under SC decoding without and with (w/) bit-flipping, and with partial rewinding (PR). 'all' and 'add' indicate averaging over all the decoding iterations and averaging only over additional iterations, respectively.}
    \label{fig:cplx_pr_flip3}
    \vspace{-5pt}
\end{figure}

Similarly, we can show a significant reduction in the complexity of the additional attempts in the SC-flip decoding algorithm. Fig. \ref{fig:cplx_pr_flip1}, \ref{fig:cplx_pr_flip2}, and \ref{fig:cplx_pr_flip3} illustrate the reduction in the node visits on average for CRC-polar codes of length $N=512$ at rates $R=1/4,1/2,3/4$. The metric used in the SC-flip implementation is similar to the one in \cite{afisiadis} as our purpose in this work is not the performance of SC-flip but to show the reduction in the complexity. Hence, a similar result can be obtained by applying the partial rewind on any variant of the SC-flip decoder.  As can be seen, the FER remains unchanged by partial rewind, while the additional decoding attempts are performed with significantly lower node visits on average. This reduction increases at high SNR regimes as  the targeted positions for bit-flipping become more accurate and their number decreases. The main contribution to this decrease is related to \ref{eq:j_p_t} where $j_p=j_p(t)$ in the fewer additional attempts, mostly one attempt.

\begin{figure}
    \centering
    \includegraphics[width=1\columnwidth]{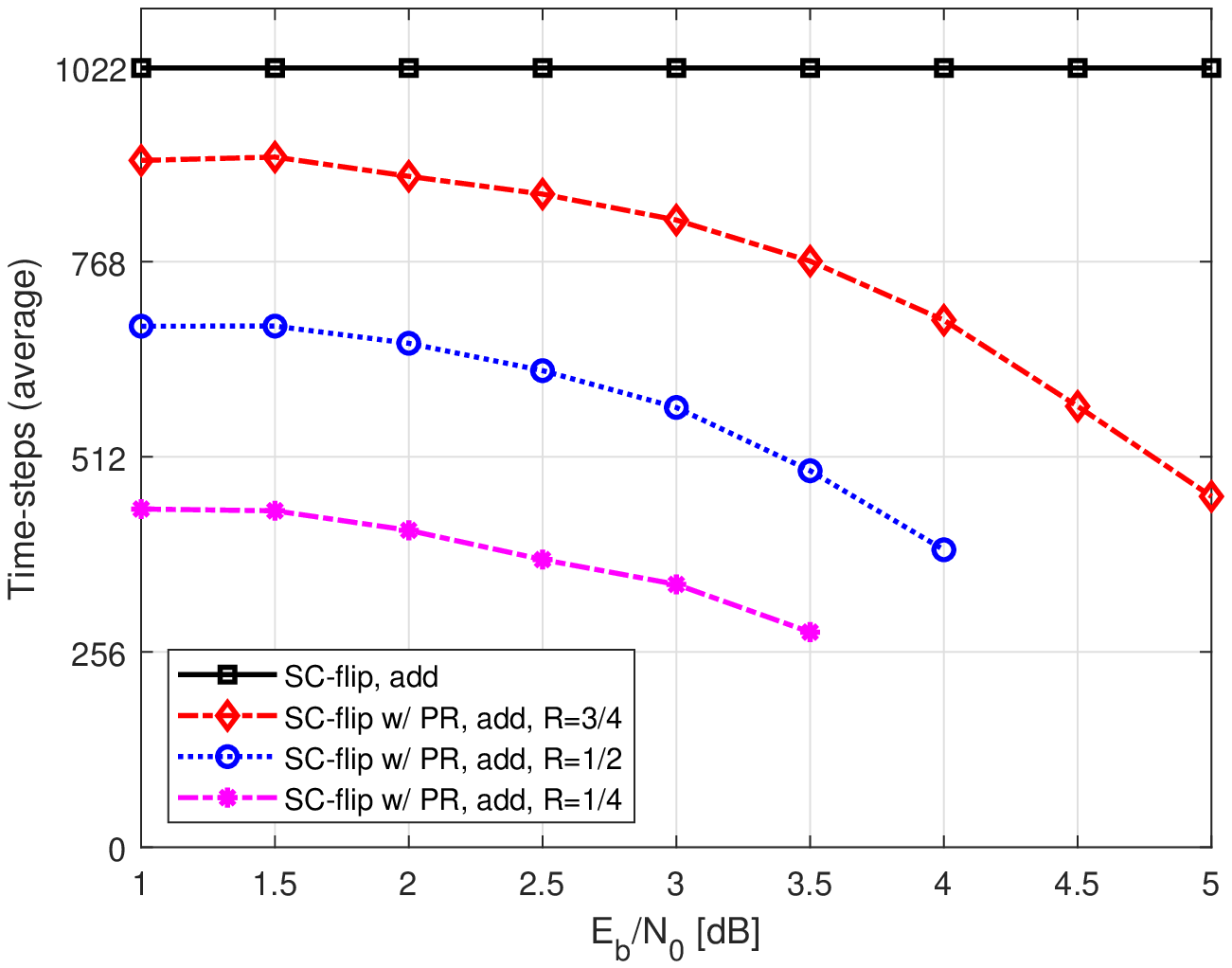}
    \caption{ }
    \label{fig:cplx_pr_flip_add}
    \vspace{-5pt}
\end{figure}

Fig. \ref{fig:cplx_pr_flip_add} compares the time complexity at 
at rates $R=1/4,1/2,3/4$. By recalling Remark \ref{rmk:nfz_bits_distrib}, one can observe that at low code rates, $(N-1)-j_p$ on average decreases significantly comparing with high rates, therefore, we expect to visit a fewer nodes in the additional decoding attempts and consequently the time complexity reduces more than high code rates. Similar to node visits, this reduction increases at high SNR regimes as  the targeted positions for bit-flipping become more accurate and their number decreases. Note that the average time complexity over additional iterations does not depend on the code rate if we don't use partial rewinding as we start redecoding from bit 0 for any code rate.

\section{CONCLUSION}
When decoding fails in the first decoding attempt, a partial rewind of the SC process for additional attempts is needed in the memory-efficient SC-based decoders. In this paper, an efficient partial rewinding approach based on the properties of the SC algorithm is proposed. This approach  relies on the properties of the SC process and its updating schedule.  Then, this scheme is adapted to multiple rewinds, and to SC list decoding, where there exists more than one path comparing with SC decoding. The numerical results show a significant reduction in the average time and computational complexity of additional decoding attempts in the SC-flip decoding and SC list decoding under the shifted-pruning scheme while the performance remains the same.

\addtolength{\textheight}{-12cm}   






\end{document}